\documentclass[a4paper]{jpconf}
\usepackage{graphicx}
\usepackage{citesort}
\begin{document}

\title{Deconfinement to Quark Matter in Magnetars}

\author{Veronica Dexheimer}
\address{UFSC, Florianopolis, BR}
\address{Gettysburg College, Gettysburg, USA}
\ead{vantoche@gettysburg.edu}

\author{Rodrigo Negreiros}
\address{Instituto de Fisica, UFF, Niteroi, BR}

\author{Stefan Schramm}
\address{FIAS - Johann Wolfgang Goethe University, Frankfurt, DE}

\begin{abstract}
We model magnetars as hybrid stars, which have a core of quark matter surrounded by hadronic matter. For this purpose, we use an extended version of the SU(3) non-linear realization of the sigma model in which the degrees of freedom change naturally from hadrons to quarks as the temperature/density increases. The presence of a variable magnetic field allows us to study in detail the influence of Landau quantization and the anomalous magnetic moment on the particle population of the star, more precisely on particles with different spin projections. This allows us to calculate the polarization of the system throughout different phases of the star, hadronic, quark and also a mixed phase.
\end{abstract}

\section{Introduction and Model Description}

Magnetars are compact stars that have extremely high magnetic fields. The highest magnetic field observed on the surface of a star is on the order of $10^{15}$ G. The highest possible magnetic field in the center of stars, on the other hand, can only be estimated through models, even when applying the virial theorem. Some results indicate limiting magnetic fields ranging between $B=10^{18}-10^{20}$ G ~\cite{Bocquet:1995je,Cardall:2000bs,1991ApJ...383..745L,%
Chakrabarty:1997ef,Bandyopadhyay:1997kh,Broderick:2001qw,Ferrer:2010wz,Malheiro:2004sb}. Here we assume that magnetars are not necessarily composed of hadronic matter and describe them using a model that contains both hadronic and quark degrees of freedom.

There is no first principles solution for QCD in both the hadronic and quark phases at finite temperature and density. In order to model the deconfinement transition, we use a model that agrees with nuclear saturation properties and reproduces reasonable hyperon optical potentials in the low energy limit and incorporates known QCD properties in the high energy limit. In order to constrain the model at intermediate energies, we compare our model predictions with lattice results such as: a first order phase transition and a pressure functional P(T) similar to Ref. \cite{Roessner:2006xn}  at  $\mu=0$ for pure gauge theory, a crossover at vanishing chemical potential with a transition temperature determined as the peak of the change of the chiral condensate and the Polyakov loop, and the location of the critical end-point \cite{Fodor:2004nz,Aoki:2006we}.

The Lagrangian density of the non-linear sigma model (shown in Ref.~\cite{Dexheimer:2009hi}) represents the interactions between baryons (and quarks) and vector and scalar mesons, the self interactions of scalar and vector mesons and includes an explicit chiral symmetry breaking term which is responsible for producing the masses of the pseudo-scalar mesons. The mesons are treated as classical fields within the mean-field approximation. Finite-temperature calculations include a heat bath of hadronic and quark quasiparticles within the grand canonical ensemble.

\begin{figure}[t]
\begin{minipage}{17.8pc}
\includegraphics[width=17.8pc]{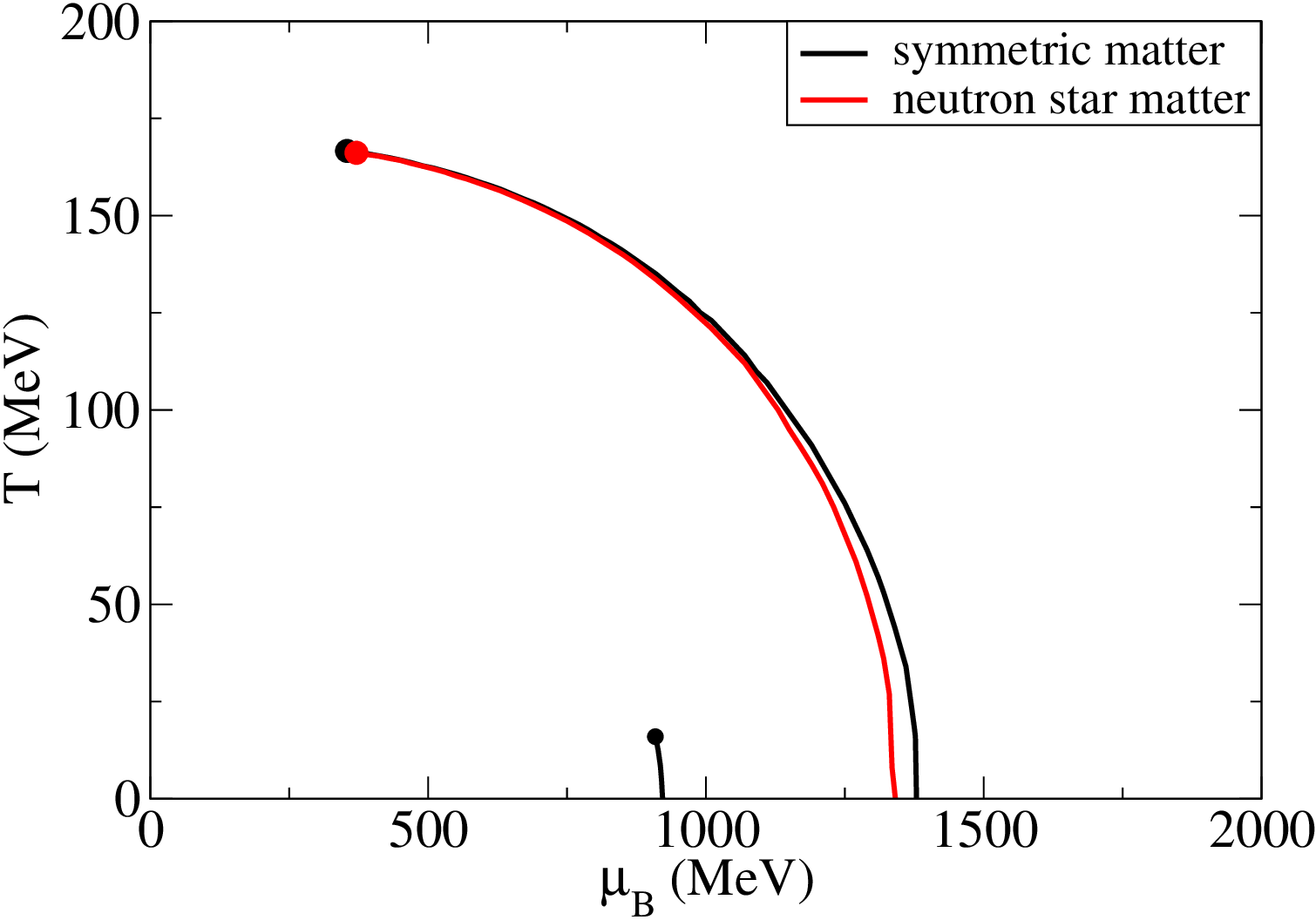}
\caption{\label{phase} (Color online) QCD Phase Diagram - Temperature as a function of baryon chemical potential showing first-order phase transitions. The dots represent critical points.}
\end{minipage}\hspace{2pc}%
\begin{minipage}{17.8pc}
\includegraphics[width=17.4pc]{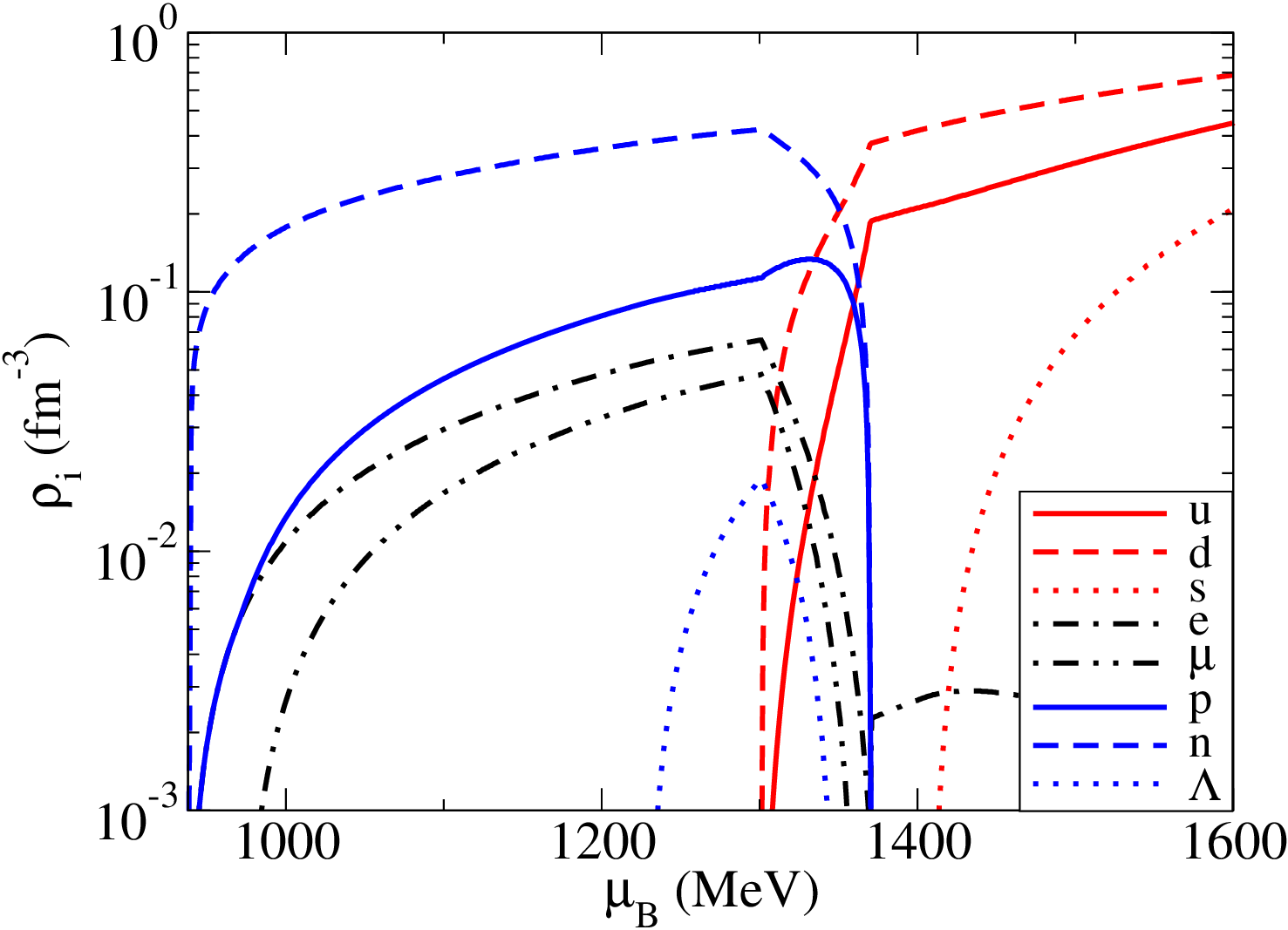}
\caption{\label{popnob} (Color online) Particle densities as a function of baryon chemical potential assuming global charge neutrality and chemical equilibrium at T=0. Quark densities are divided by 3.}
\end{minipage} 
\end{figure}

Within our approach, the hadrons (whole baryon octet) are replaced by quarks (up, down, strange) at high densities and/or temperatures. This happens as the effective masses of the hadrons increase and the effective masses of the quarks decrease within these limits. The aforementioned model (see Ref.~\cite{Dexheimer:2009hi} for details) is an extended version of the SU(3) non-linear realization of the sigma model. Changes in the order parameters of the model $\sigma$ and $\Phi$ signal chiral symmetry restoration and quark deconfinement, respectively. The potential for $\Phi$ is an extension of the Polyakov loop potential \cite{Roessner:2006xn}  modified to also depend on baryon chemical potential. In this way our model is able to describe the entire QCD phase diagram, even at zero temperature. Fig.~\ref{phase} shows that the model is in good agreement with lattice QCD constraints and that it reproduces the liquid-gas phase transition for symmetric matter. In this figure we also show results for neutron star matter, which is charge neutral and in chemical equilibrium. The phase transitions at low temperatures and high densities are of first order, whereas at high temperatures and low densities the model exhibits smooth crossovers. 

The SU(3) non-linear realization of the sigma model and its extension (that also contains quarks) have been successful in reproducing nuclear matter properties  \cite{Papazoglou:1998vr}, heavy ion collision data \cite{Steinheimer:2009nn}, compact star and proto-neutron star properties \cite{Dexheimer:2008ax,Negreiros:2010hk,Dexheimer:2011pz}.
As compact stars have temperatures of the order of $1$ MeV, we can safely set their temperature to zero. As already mentioned, for star calculations we have to take into account charge neutrality and chemical equilibrium. Here we assume that the surface tension between the hadronic and quark phases is small \cite{Pinto:2012aq} and allow charge neutrality to be global (only the combination of both phases sum up to zero charge). As a consequence, a mixed phase appears in the star. This can be seen in Fig.~\ref{popnob}, which also shows that hyperons are almost completely suppressed by the appearance of the quarks.

We include in the model a magnetic field  in the z-direction that has varying magnitude. This is a more realistic approach than considering a constant magnetic field throughout the star and can prevent the creation of hydrodynamical instabilities due to pressure anisotropy \cite{Chaichian:1999gd,PerezMartinez:2005av,PerezMartinez:2007kw,Huang:2009ue,Paulucci:2010uj}. This happens because, in our approach, the magnetic field only becomes extremely high in the center of the star, where the matter pressure is also high (see Ref.~\cite{Dexheimer:2012mk} for more details).  More precisely, we assume an effective magnetic field $B^*$ that increases with chemical potential, running from a surface value $B_{surf}=69.25$ MeV$^2=10^{15}$ G (when $\mu_B=938$ MeV) to different central values $B_c$ at large values of baryon chemical potential following \cite{Dexheimer:2011pz}
\begin{equation}
B^*(\mu_B)=B_{surf}+B_c[1-e^{b\frac{(\mu_B-938)^a}{938}}],
\end{equation}
with $a=2.5$, $b=-4.08\times10^{-4}$ and $\mu_B$ given in MeV. As can be seen in Fig.~\ref{beff}, the values of the effective magnetic field only approach $B_c$ at very high baryon chemical potentials and, in practice, only about $70\%$ of $B_c$ can be reached inside stars. The use of an explicit dependence of B on the baryon chemical potential instead of on density was chosen to prevent discontinuities in the magnetic field at the phase transition, where the baryon density is discontinuous. The constants $a$ and $b$ and the form of the $B^*$ expression were chosen to reproduce (in the absence of quarks) the effective magnetic field curve as a function of density from Refs.~\cite{Bandyopadhyay:1997kh,Mao:2001fq,Rabhi:2009ih}.

\section{Results and Conclusions}

\begin{figure}[t]
\begin{minipage}{17.8pc}
\includegraphics[width=18.8pc]{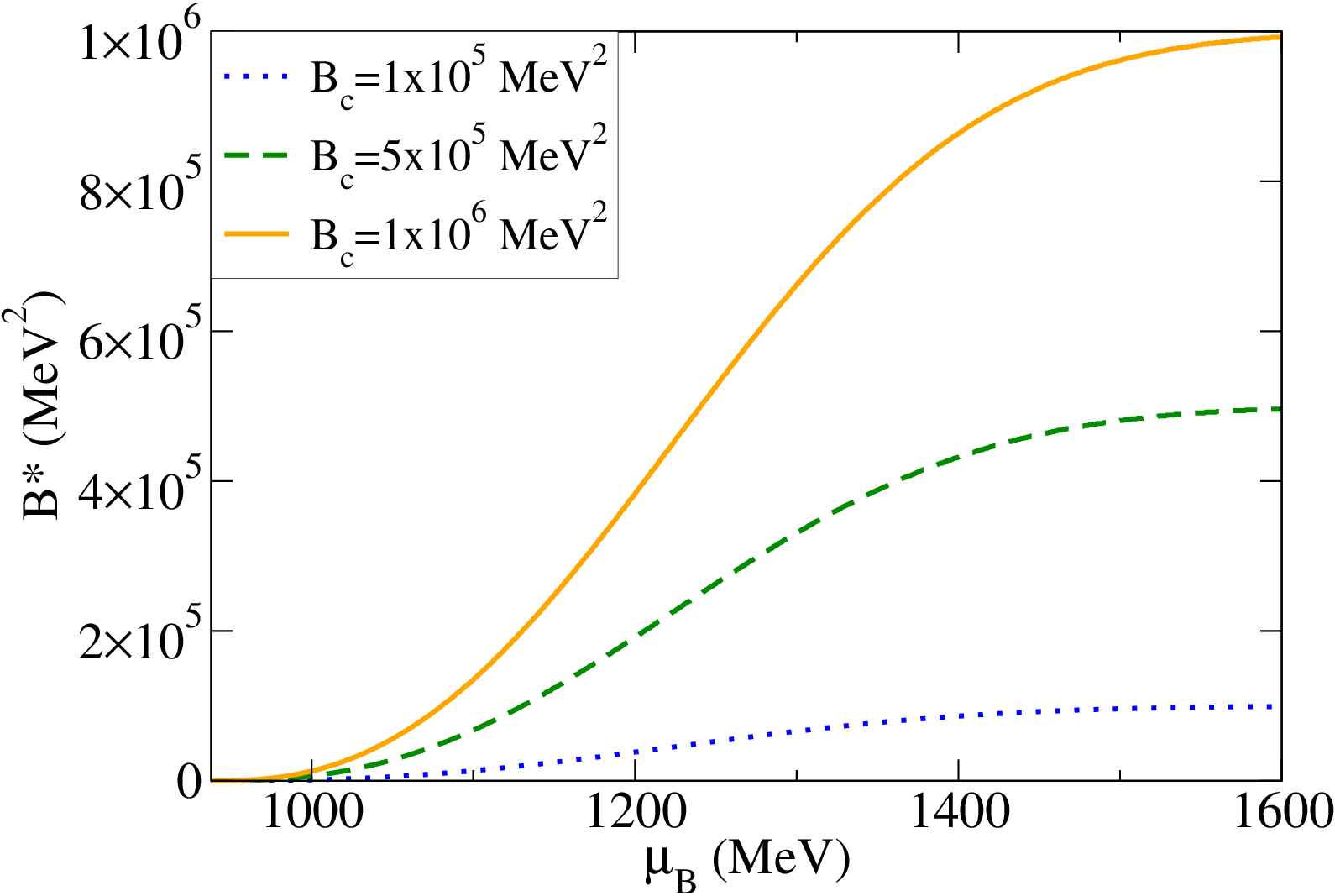}
\caption{\label{beff} (Color online) Effective magnetic field as a function of baryon chemical potential shown for different central magnetic fields. Recall that using Gaussian natural units $1$ MeV$^2=1.44\times10^{13}$ G. \\ \vspace{.5cm} }
\end{minipage}\hspace{2pc}%
\begin{minipage}{17.8pc}
\includegraphics[width=17.8pc]{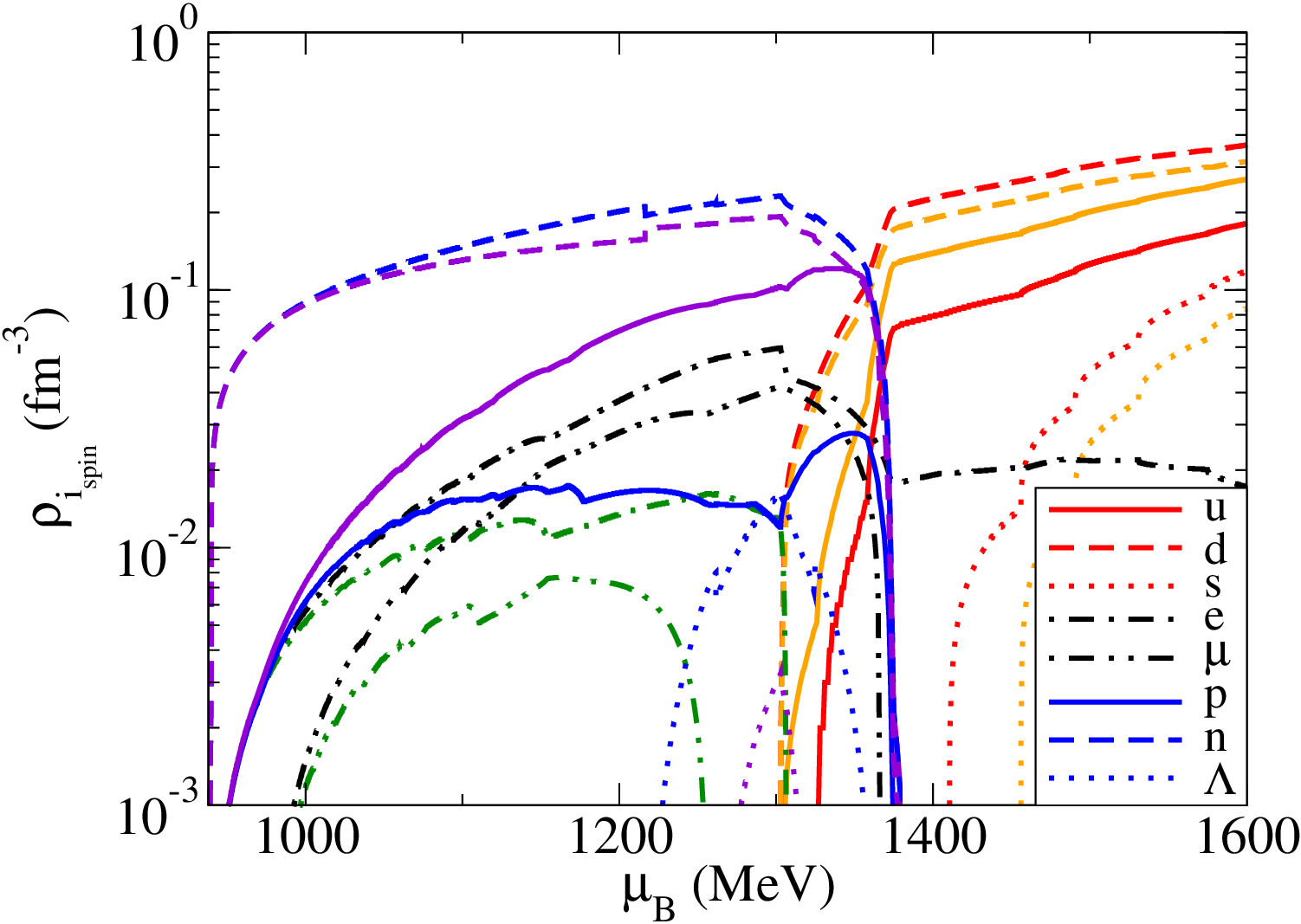}
\caption{\label{popspin} (Color online) Particle densities as a function of baryon chemical potential for $B_c=5\times10^5$ MeV$^2=7.22\times10^{18}$ G including AMM. Black, blue and red stand for negative, while green, purple and orange stand for positive spin projections. Quark densities are divided by 3.}
\end{minipage} 
\end{figure}

The magnetic field in the z-direction forces the eigenstates in the x and y directions of charged particles to be quantized into Landau levels. The energy levels of all baryons are further split due to the alignment/anti-alignment of their spins with the magnetic field (anomalous magnetic moment effect, AMM). But even when the AMM is not taken into account, like in the quark phase in our model, only one of the spin projections contributes to the zeroth Landau level, creating a spin projection asymmetry in the system. In this work, we focus on the analysis of magnetic field effects on the chemical composition of the neutron star, the total spin polarization and the magnetization of the system. Studies of magnetic field effects on compact star observables can be found in Refs.~\cite{Dexheimer:2011pz,Menezes:2009uc,Menezes:2008qt,2011PhRvC..83f5805A}.

The particle population is shown in Fig.~\ref{popspin} when a central magnetic field $B_c=5\times10^5$ MeV$^2=7.22\times10^{18}$ G with AMM is considered. The ``wiggles'' in the charged particle densities mark the $\mu_B$ values, for which their Fermi energies cross the discrete threshold of a Landau level. The charged particle population is enhanced due to $B$, as their chemical potentials increase. Although the AMM is known to make the EOS stiffer, it does not have a very significant effect in the particle population. This fact can be easily understood in terms of polarization, when, instead of looking at the total particle density (sum of spin up and down particle densities) for each species, we look at the spin up/spin down particle densities separately. In this case some of these particles are enhanced while others are suppressed (see Ref.~\cite{Dexheimer:2012qk} for details).

\begin{figure}[t]
\begin{minipage}{17.8pc}
\includegraphics[width=17.4pc]{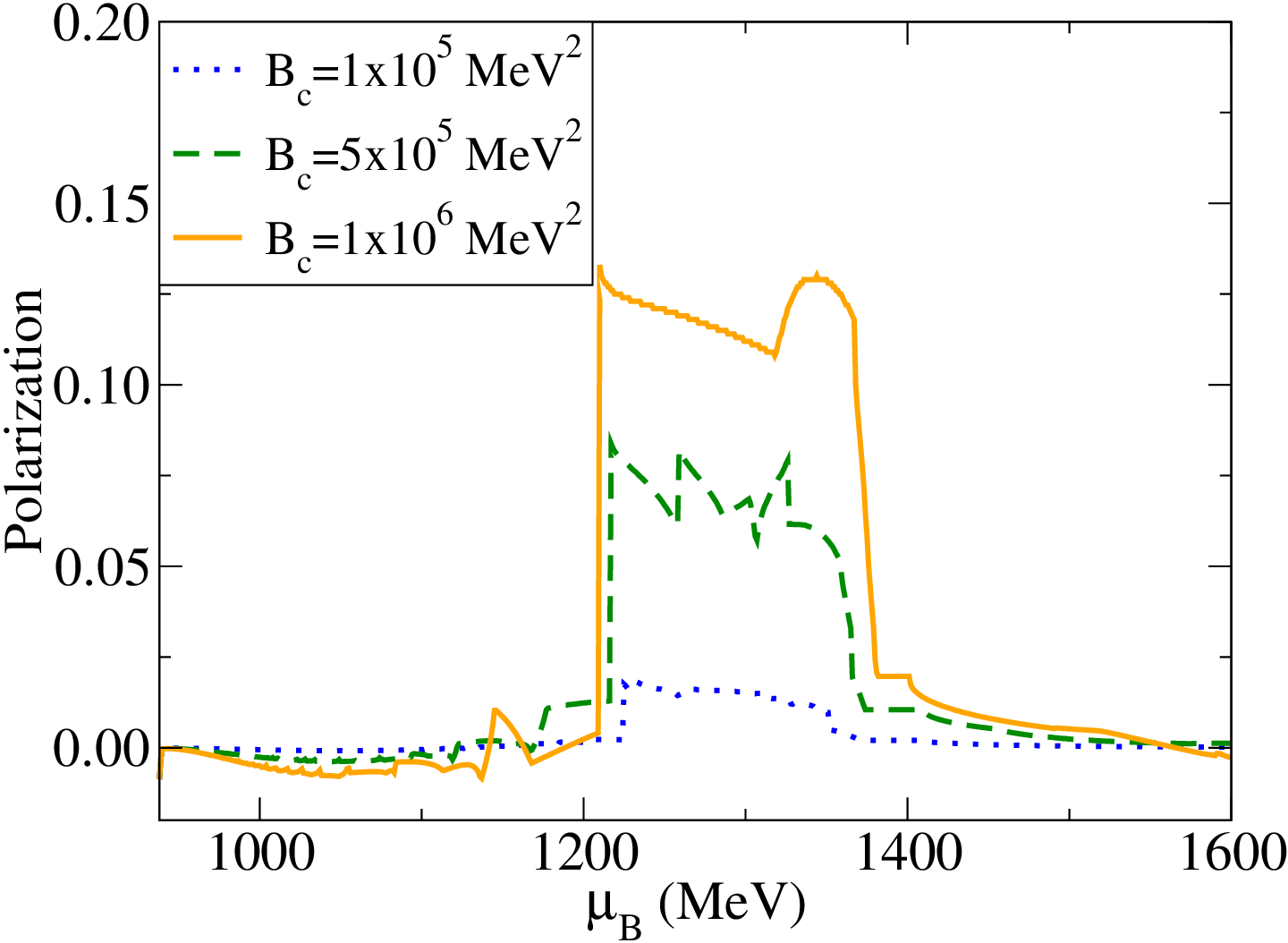}
\caption{\label{polar} (Color online) Total spin polarization as a function of baryon chemical potential for different central magnetic fields (with $1$ MeV$^2=1.44\times10^{13}$ G) including AMM. \vspace{.1000 cm}} 
\end{minipage}\hspace{2pc}%
\begin{minipage}{17.8pc}
\includegraphics[width=17.8pc]{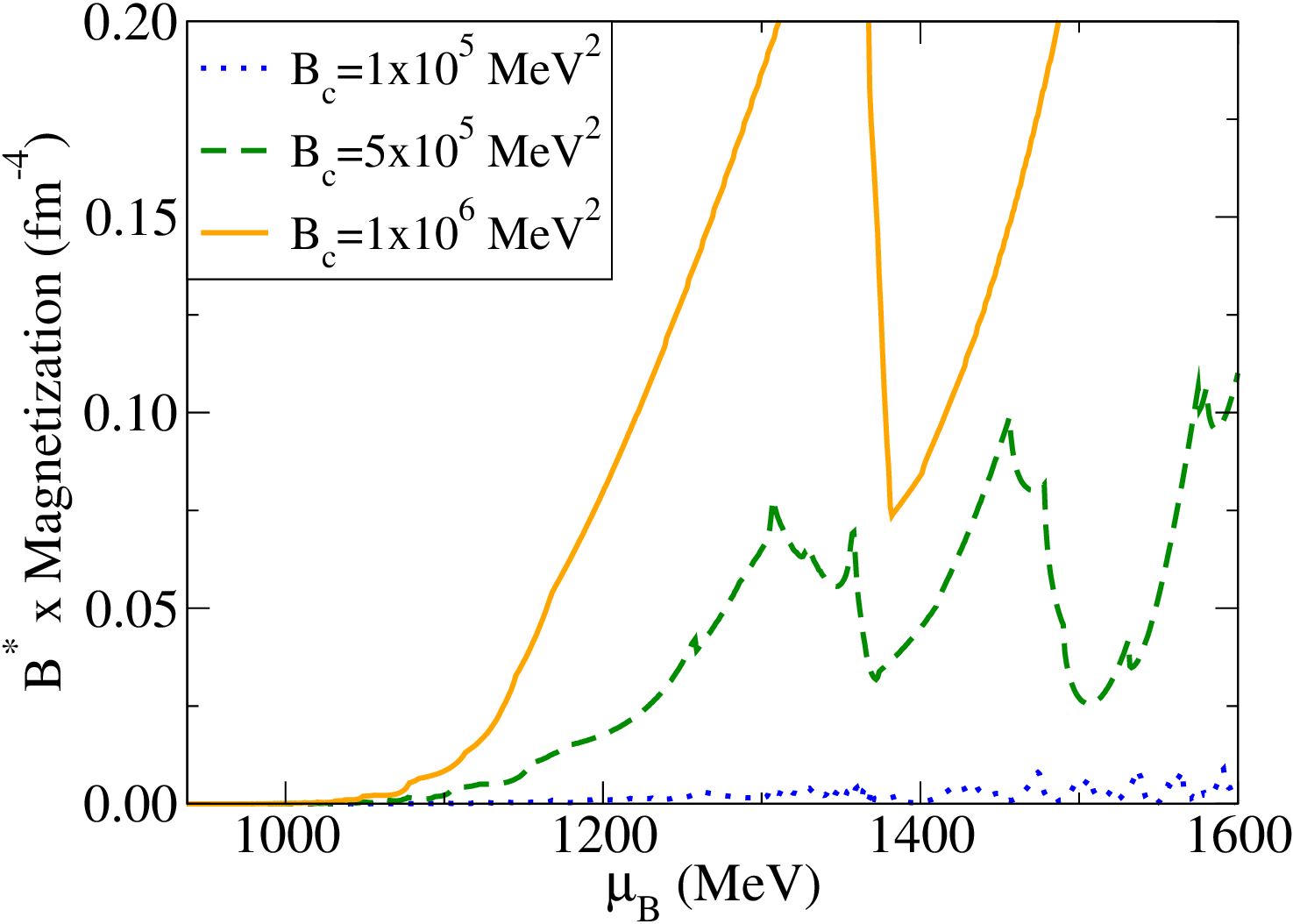}
\caption{\label{mag2} (Color online) Effective magnetic field times magnetization as a function of baryon chemical potential for different central magnetic fields (with $1$ MeV$^2=1.44\times10^{13}$ G) including AMM.}
\end{minipage} 
\end{figure}

The total polarization of the system is defined as 
\begin{equation}
\frac{\sum_i Q_{Bi} (\rho_{+i}-\rho_{-i})}{\sum_i Q_{Bi} (\rho_{+i}+\rho_{-i})},
\end{equation}
where $Q_{Bi}$ stands for the baryon number of each species ($3$ for baryons, $1$ for quarks and $0$ for leptons). We can see in Fig.~\ref{polar} that the total polarization of the system (taking into account hadrons and quarks) is larger for larger magnetic fields and increases with chemical potential until the quarks appear. After this point the polarization decreases. This happens because the AMM for the quarks is not included in our calculations, as it is not fully understood for these particles. In this case, the spin asymmetry comes exclusively from the spin asymmetric contribution to the zeroth Landau level. For an example of a complete study of the polarization without the AMM see Ref.~\cite{Isayev:2012sv}. The ``wiggles'' in Fig.~\ref{polar} mark again when the Fermi energies of the charged particles cross the discrete threshold of a Landau level.

Finally, we turn our attention to the calculation of the magnetization of the system. It is important to notice that in the case with the AMM, not only the magnetization of the charged particles has to be included, but also the magnetization of the uncharged particles. For details on the calculation of the magnetization including the AMM see Ref.~\cite{Strickland:2012vu}. Fig.~\ref{mag2} shows that the magnetization is larger for larger magnetic fields and it increases with baryon chemical potential. As in Figs.~\ref{popspin} and \ref{polar}, the "wiggles" mark when the Fermi energies of the charged particles cross the discrete threshold of a Landau level. The magnetization of the system is a very important quantity as it relates with the pressure anisotropy of the system \cite{Chaichian:1999gd,PerezMartinez:2005av,PerezMartinez:2007kw,Huang:2009ue,Paulucci:2010uj}.

We have shown in this work some possible effects of strong magnetic fields in hybrid stars. The presence of different hadronic and quark degrees of freedom makes this quark-hadron sigma model an ideal tool for such an analysis in the different possible phases of the star. More specifically, we analyzed the effects of strong chemical potential dependent magnetic fields on particles with different spin projections, their total polarization and magnetization.  

\section*{Acknowledgments}
V. D. acknowledges support from CNPq (National Counsel of Technological and Scientific Development - Brazil).

\section*{References}
\bibliography{Paper}

\providecommand{\newblock}{}
\begin{thebibliography}{10}
\expandafter\ifx\csname url\endcsname\relax
  \def\url#1{{\tt #1}}\fi
\expandafter\ifx\csname urlprefix\endcsname\relax\def\urlprefix{URL }\fi
\providecommand{\eprint}[2][]{\url{#2}}

\bibitem{Bocquet:1995je}
Bocquet M, Bonazzola S, Gourgoulhon E and Novak J 1995 {\em
  Astron.Astrophys.\/} {\bf 301} 757 (\textit{Preprint} \eprint{gr-qc/9503044})

\bibitem{Cardall:2000bs}
Cardall C~Y, Prakash M and Lattimer J~M 2001 {\em Astrophys.J.\/} {\bf 554}
  322--339 (\textit{Preprint} \eprint{astro-ph/0011148})

\bibitem{1991ApJ...383..745L}
{Lai} D and {Shapiro} S~L 1991 {\em Astrophys. J.\/} {\bf 383} 745--751

\bibitem{Chakrabarty:1997ef}
Chakrabarty S, Bandyopadhyay D and Pal S 1997 {\em Phys.Rev.Lett.\/} {\bf 78}
  2898--2901 (\textit{Preprint} \eprint{astro-ph/9703034})

\bibitem{Bandyopadhyay:1997kh}
Bandyopadhyay D, Chakrabarty S and Pal S 1997 {\em Phys. Rev. Lett.\/} {\bf 79}
  2176--2179 (\textit{Preprint} \eprint{astro-ph/9703066})

\bibitem{Broderick:2001qw}
Broderick A~E, Prakash M and Lattimer J~M 2002 {\em Phys. Lett.\/} {\bf B531}
  167--174 (\textit{Preprint} \eprint{astro-ph/0111516})

\bibitem{Ferrer:2010wz}
Ferrer E~J, de~la Incera V, Keith J~P, Portillo I and Springsteen P~P 2010 {\em
  Phys. Rev.\/} {\bf C82} 065802 (\textit{Preprint} \eprint{1009.3521})

\bibitem{Malheiro:2004sb}
Malheiro M, Ray S, Mosquera~Cuesta H~J and Dey J 2007 {\em Int.J.Mod.Phys.\/}
  {\bf D16} 489--499 (\textit{Preprint} \eprint{astro-ph/0411675})

\bibitem{Roessner:2006xn}
Roessner S, Ratti C and Weise W 2007 {\em Phys.Rev.\/} {\bf D75} 034007
  (\textit{Preprint} \eprint{hep-ph/0609281})

\bibitem{Fodor:2004nz}
Fodor Z and Katz S 2004 {\em JHEP\/} {\bf 0404} 050 (\textit{Preprint}
  \eprint{hep-lat/0402006})

\bibitem{Aoki:2006we}
Aoki Y, Endrodi G, Fodor Z, Katz S and Szabo K 2006 {\em Nature\/} {\bf 443}
  675--678 (\textit{Preprint} \eprint{hep-lat/0611014})

\bibitem{Dexheimer:2009hi}
Dexheimer V~A and Schramm S 2010 {\em Phys. Rev.\/} {\bf C81} 045201
  (\textit{Preprint} \eprint{0901.1748})

\bibitem{Papazoglou:1998vr}
Papazoglou P {\em et~al.\/} 1999 {\em Phys. Rev.\/} {\bf C59} 411--427
  (\textit{Preprint} \eprint{nucl-th/9806087})

\bibitem{Steinheimer:2009nn}
Steinheimer J, Dexheimer V, Petersen H, Bleicher M, Schramm S {\em et~al.\/}
  2010 {\em Phys.Rev.\/} {\bf C81} 044913 (\textit{Preprint}
  \eprint{0905.3099})

\bibitem{Dexheimer:2008ax}
Dexheimer V and Schramm S 2008 {\em Astrophys. J.\/} {\bf 683} 943--948
  (\textit{Preprint} \eprint{0802.1999})

\bibitem{Negreiros:2010hk}
Negreiros R, Dexheimer V and Schramm S 2010 {\em Phys.Rev.\/} {\bf C82} 035803
  (\textit{Preprint} \eprint{1006.0380})

\bibitem{Dexheimer:2011pz}
Dexheimer V, Negreiros R and Schramm S 2012 {\em Eur.Phys.J.\/} {\bf A48} 189
  (\textit{Preprint} \eprint{1108.4479})

\bibitem{Pinto:2012aq}
Pinto M~B, Koch V and Randrup J 2012 {\em Phys.Rev.\/} {\bf C86} 025203
  (\textit{Preprint} \eprint{1207.5186})

\bibitem{Chaichian:1999gd}
Chaichian M, Masood S, Montonen C, Perez~Martinez A and Perez~Rojas H 2000 {\em
  Phys.Rev.Lett.\/} {\bf 84} 5261--5264 (\textit{Preprint}
  \eprint{hep-ph/9911218})

\bibitem{PerezMartinez:2005av}
Perez~Martinez A, Perez~Rojas H, Mosquera~Cuesta H~J, Boligan M and Orsaria M~G
  2005 {\em Int. J. Mod. Phys.\/} {\bf D14} 1959 (\textit{Preprint}
  \eprint{astro-ph/0506256})

\bibitem{PerezMartinez:2007kw}
Perez~Martinez A, Perez~Rojas H and Mosquera~Cuesta H 2008 {\em Int. J. Mod.
  Phys.\/} {\bf D17} 2107--2123 (\textit{Preprint} \eprint{0711.0975})

\bibitem{Huang:2009ue}
Huang X~G, Huang M, Rischke D~H and Sedrakian A 2010 {\em Phys.Rev.\/} {\bf
  D81} 045015 (\textit{Preprint} \eprint{0910.3633})

\bibitem{Paulucci:2010uj}
Paulucci L, Ferrer E~J, de~la Incera V and Horvath J~E 2011 {\em Phys. Rev.\/}
  {\bf D83} 043009 (\textit{Preprint} \eprint{1010.3041})

\bibitem{Dexheimer:2012mk}
Dexheimer V, Menezes D and Strickland M 2012  (\textit{Preprint}
  \eprint{1210.4526})

\bibitem{Mao:2001fq}
Mao G~J, Iwamoto A and Li Z~X 2003 {\em Chin. J. Astron. Astrophys.\/} {\bf 3}
  359--374 (\textit{Preprint} \eprint{astro-ph/0109221})

\bibitem{Rabhi:2009ih}
Rabhi A, Pais H, Panda P~K and Providencia C 2009 {\em J. Phys.\/} {\bf G36}
  115204 (\textit{Preprint} \eprint{0909.1114})

\bibitem{Menezes:2009uc}
Menezes D, Benghi~Pinto M, Avancini S and Providencia C 2009 {\em Phys.Rev.\/}
  {\bf C80} 065805 (\textit{Preprint} \eprint{0907.2607})

\bibitem{Menezes:2008qt}
Menezes D, Benghi~Pinto M, Avancini S, Perez~Martinez A and Providencia C 2009
  {\em Phys.Rev.\/} {\bf C79} 035807 (\textit{Preprint} \eprint{0811.3361})

\bibitem{2011PhRvC..83f5805A}
{Avancini} S~S, {Menezes} D~P and {Providencia} C 2011 {\em "Phys. Rev. C"\/}
  {\bf 83} 065805

\bibitem{Dexheimer:2012qk}
Dexheimer V, Negreiros R, Schramm S and Hempel M 2012  (\textit{Preprint}
  \eprint{1208.1320})

\bibitem{Isayev:2012sv}
Isayev A and Yang J 2013 {\em J.Phys.\/} {\bf G40} 035105 (\textit{Preprint}
  \eprint{1210.3322})

\bibitem{Strickland:2012vu}
Strickland M, Dexheimer V and Menezes D 2012 {\em Phys.Rev.\/} {\bf D86} 125032
  (\textit{Preprint} \eprint{1209.3276})

\end{thebibliography}
\end{document}